# Statistical properties of random clique networks

Yi-Min Ding [1,2], Jun Meng [3], Jing-Fang Fan [2], Fang-Fu Ye [3,4,*], Xiao-Song Chen [2,4,‡]

[1] Faculty of Physics and Electronics, Hubei University, Wuhan 430062, China
[2] CAS Key Laboratory of Theoretical Physics, Institute of Theoretical Physics, Chinese Academy of Sciences, P.O. Box 2735, Beijing 100190, China
[3] Beijing National Laboratory for Condensed Matter Physics, CAS Key Laboratory of Soft Matter Physics, Institute of Physics, Chinese Academy of Sciences, Beijing 100190, China,
[4] School of Physical Sciences, University of Chinese Academy of Sciences, Beijing 100049, China
Corresponding authors. E-mail: ‡chenxs@itp.ac.cn, *fye@iphy.ac.cn

In this paper, a random clique network model to mimic the large clustering coefficient and the modular structure that exist in many real complex networks, such as social networks, artificial networks, and protein interaction networks, is introduced by combining the random selection rule of the Erdös and Rényi (ER) model and the concept of cliques. We find that random clique networks having a small average degree differ from the ER network in that they have a large clustering coefficient and a power law clustering spectrum, while networks having a high average degree have similar properties as the ER model. In addition, we find that the relation between the clustering coefficient and the average degree shows a non-monotonic behavior and that the degree distributions can be fit by multiple Poisson curves; we explain the origin of such novel behaviors and degree distributions.

## I. INTRODUCTION

In the past decades, various models of complex networks have been proposed, including the Erdös and Rényi (ER) model [1], small-world model [2], and scale-free model [3, 4]. One common feature of these models is that they are built by adding edges. After setting certain rules as to how the edges are added, these models are able to reproduce some important features of complex networks in the real world, such as small-world and scale-free features [5–8]. However, there are still some features that cannot be explained by these models. One important example is the hierarchical structure of real networks [9–13], e.g., biological cells include many large bio-molecules, and these molecules are composed of various atoms.

In recent years, the concept of cliques or clique networks has been proposed to understand the hierarchical structure of real complex networks [14–26]. A clique of size $m$ is a completely connected unit in a network that has $m$ nodes and $m(m − 1)/2$ edges. In 2005, Palla et al. introduced the concept of clique percolation and used the clique percolation method to investigate overlapping graph communities [15, 16]. Takemoto et al. constructed evolving clique networks by merging individual cliques, and thus, obtained hierarchical scale-free networks with high clustering coefficients [17]. Some scholars introduced the concept of hypergraph to explain the hierarchical structure of real complex networks [19-21], In 2014, Ding et al. built hybrid evolving clique networks to explain the inhomogeneous and/or homogeneous features of real hierarchical complex systems [23]. Recently, Slater et al. presented an empirical study on real world networks and found that medium-sized cliques are more common in real world networks than triangles [24]; these results suggest that the concept of clique is a central organizing principle of many real world networks.

In the meanwhile, the classical ER model [1] still remains of great interest because it can serve as a test bed for checking various new ideas of complex networks. Thus, in this paper, we propose combining the ER selection rule with the concept of clique and construct random clique networks to investigate the consequences of such a combination. In Sec. II, we first describe how to build such networks, and in the following sections, various properties of these networks will be presented.

## II. RANDOM CLIQUE NETWORK MODEL

Inspired by studies on the modular structure of complex networks and those on the ER random network, we introduce a random clique network (RCN) model in this paper. Start from $N$ isolated nodes and connect every $m$ node with probability $p$ to form a complete subgraph of $m$-clique. Fig. 1 shows two network examples with $N = 10$ and $m = 3$. When $p$ is small [see Fig. 1, where $p = 0.03$], the cliques are not completely connected to each other, i.e., there exist some isolated cliques or clique clusters; as $p$ increases, all the cliques connect to each other and form a big clique cluster.

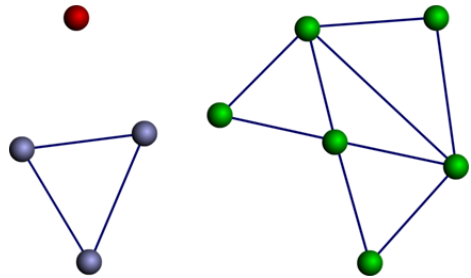

FIG. 1: (Color online) Sketches of the random clique graph with $N = 10, p = 0.03$.

The number of small clique clusters can be calculated explicitly. Here, we illustrate the calculation for the case where $m = 3$ (the calculation for $m = 4$ can be carried out in a similar way), and the calculation results will be used in the sections below. Following the definition of the connection probability $p$, we can obtain the total number of cliques as follows

$$N^T = p\binom{N}{3}. \quad (1)$$

We then compute the total number of pairs of cliques that share one node, which is equal to the product of $p^2$ and the possible ways of forming such configurations from $N$ nodes, i.e.,

$$N_{NP}^T = p^2 \binom{N}{5}\binom{5}{3}\binom{3}{1}\binom{3}{2}\frac{1}{2!} \approx \frac{1}{8}N^5 p^2, \quad (2)$$

where we first select five nodes from $N$ nodes, and then select any three from these five nodes to form a clique. We then choose the node to be shared by two cliques from these three nodes. The factorial in the above equation represents the symmetry of these two cliques. Similarly, we can obtain the total number of pairs of cliques that share one edge,

$$N_{EP}^T = p^2 \binom{N}{4}\binom{4}{3}\binom{3}{2}\frac{1}{2!} \approx \frac{1}{4}N^4 p^2. \quad (3)$$

The different categories of configurations that can be formed by three connected cliques are listed in Fig. 2. We give in the following the total number of each category:

$$N_1 = p^3 \binom{N}{7}\binom{7}{3}\binom{3}{1}\binom{4}{2}\binom{2}{1}\binom{2}{2}\frac{1}{2!} \approx \frac{1}{8}N^7 p^3, \quad (4)$$

$$N_2 = p^3 \binom{N}{7}\binom{7}{3}\binom{3}{1}\binom{4}{2}\binom{2}{2}\frac{1}{3!} \approx \frac{1}{48}N^7 p^3, \quad (5)$$

$$N_3 = p^3 \binom{N}{6}\binom{6}{3}\binom{3}{1}\binom{3}{2}\binom{1}{1}\frac{1}{1!} \approx \frac{1}{2}N^6 p^3, \quad (6)$$

$$N_4 = p^3 \binom{N}{6}\binom{6}{3}\binom{3}{1}\binom{3}{2}\binom{1}{1}\frac{1}{2!2!} \approx \frac{1}{8}N^6 p^3, \quad (7)$$

$$N_5 = p^3 \binom{N}{5}\binom{5}{3}\binom{3}{2}\binom{2}{1}\binom{1}{1}\frac{1}{2!} \approx \frac{1}{4}N^5 p^3. \quad (8)$$

We can see from Eqs. (6), (7) and (8) that the number of configurations with shared edges is much smaller than those with shared nodes.

Our model is identical to the ER model when $m = 2$. However, for $m > 2$, the statistical properties of RCNs differ from those of the ER random network. We will investigate such properties in the following sections. We will present both analytical and numerical results, with the latter

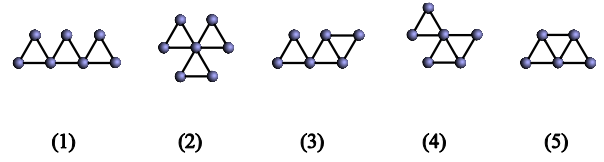

(1)  (2)  (3)  (4)  (5)

FIG. 2: (Color online) Categories of clusters formed by three cliques of size $m = 3$.

obtained by Monte Carlo simulations.

## III. DEGREE DISTRIBUTION

We first study the degree distributions of these networks. As shown in Fig. 3, the distributions can now be divided into multiple categories. For $m = 3$ [see Fig. 3(b) and (d)], there exist two categories—the even-number degrees and the odd-number degrees, with the former having much higher probabilities than the latter, and the difference between these two categories increases with an increase in the network size $N$. For $m = 4$, the distributions can be divided into three categories: (i) $k = 3n$, (ii) $k = 3n - 1$, and (iii) $k = 3n-2$, where $n$ represents a positive integer; Category I has higher probabilities than Category II, and Category II has higher probabilities than Category III.

We also perform analytical calculations to understand the above numerical results of degree distributions. We start with networks constructed with $m = 3$ cliques. The maximum number of cliques that can be potentially connected to a given node $a$ is $\beta = (N-1)(N-2)/2$. The probabilities of having exactly $n$ given cliques connected is $p^n(1-p)^{\beta-n}$. To calculate the degree of node $a$ (i.e., the number of edges connected to node $a$), we then need to know how the cliques share edges. Given that $p$ is small, we focus on two cases only: (i) there is no edge shared by cliques; (ii) there is only one edge shared by two cliques. (There exist other cases—more shared edges or more than two cliques sharing one edge. Given that $p$ is small, the probabilities for such cases to occur are small and we shall thus ignore them.) For Case I, we have degree $k = 2n$; and the corresponding probability is

$$P_{2n} = \left[\frac{1}{n!}\prod_{i=0}^{n-1}\binom{N-1-2i}{2}\right] \cdot p^n(1-p)^{\beta-n}, \quad (9)$$

where the prefactor represents the number of ways of selecting $2n$ nodes from $N-1$ nodes in $n$ steps with two nodes being selected in each step. Eq. (9) can be further simplified, and approximated by a Poisson distribution, when $p$ is small,

$$P_{2n} = \frac{(2n)!}{2^n n!}\binom{N-1}{2n} \cdot p^n(1-p)^{\beta-n} \quad (10)$$

$$\approx \frac{(2n)!}{2^n n!}\frac{\binom{N-1}{2n}}{\binom{\beta}{n}}\left(\frac{\lambda^n}{n!}\right)e^{-\lambda}, \quad (11)$$

where $\lambda = p\beta$, when $\beta \to \infty$, the above equation becomes exactly a Poisson distribution. For Case II, the degree is $k = 2n - 1$, and the number of ways of selecting $n$ cliques is the product of the number of ways of selecting the two cliques that share one edge, which is given by $(N-1)(N-2)(N-3)/2$, and that of selecting the $n-2$ non-edge-sharing cliques from the remaining $N-4$
nodes, which is given by $\Lambda = \frac{1}{(n-2)!}\prod_{i=0}^{n-3}\binom{N-4-2i}{2}$. We thus obtain the following equation:

$$P_{2n-1} = \frac{(N-1)(N-2)(N-3)\Lambda}{2} \cdot p^n(1-p)^{\beta-n}$$

$$= \frac{(2n-1)!}{2^{n-1}(n-2)!}\binom{N-1}{2n-1} \cdot p^n(1-p)^{\beta-n} \quad (12)$$

$$\approx \frac{(2n-1)!}{2^{n-2}(n-1)!}\frac{\binom{N-1}{2n}}{\binom{\beta}{n}} \cdot (\frac{\lambda^n}{n!})e^{-\lambda}, \quad (13)$$

which can also be approximated by a Poisson distribution. From Eq. (10) and Eq. (12), we find that $P_{2n}/P_{2n-1} = (N - 2n)/[2n(n-1)] \gg 1$.

The degree distribution of networks composed of $m = 4$ cliques can be similarly deduced. We consider three cases: (i) no shared edge, (ii) one shared edge, and (iii) two shared edges. The corresponding probabilities of these three cases are, respectively,

$$P_{3n} = \frac{(3n)!}{6^n n!} \cdot \binom{N-1}{3n} \cdot p^n(1-p)^{\gamma-n}, \quad (14)$$

$$P_{3n-1} = \frac{(3n-1)!}{8 \cdot 6^{n-2}(n-2)!} \cdot \binom{N-1}{3n-1} \cdot p^n(1-p)^{\gamma-n}, \text{ and} \quad (15)$$

$$P_{3n-2} = [\frac{(3n-2)!}{4 \cdot 6^{n-2}(n-2)!} + \frac{(3n-2)!}{8 \cdot 6^{n-2}(n-3)!} +$$
$$\frac{(3n-2)!}{2^7 \cdot 6^{n-4}(n-4)!}] \cdot \binom{N-1}{3n-2} \cdot p^n(1-p)^{\gamma-n}. \quad (16)$$

Here, $\gamma = (N-1)(N-2)(N-3)/6$ is the maximum number of four-cliques that can be formed with a given node $a$. The three terms in Eq. (16) represent, respectively, three edge-sharing possibilities: (i) two edges shared by two cliques, (ii) two edges shared by three cliques, and (iii) two edges shared by four cliques.

The analytical results of the degree distributions are shown in Fig.3, and they agree well with the corresponding numerical results. Given the degree distributions, the average degrees of the networks can then be calculated. We focused on the no shared edge case because the probabilities of edge sharing are relatively small. Thus, there exists a simple linear relation between the average degree and average clique number of the networks; the latter can be easily computed by using the generating function method. For $m = 3$, the clique generating function can be written as

$$G_0(z) = \sum_{n=0}^{\beta} \binom{\beta}{n} \cdot p^n (1-p)^{\beta-n} z^n = (pz + 1 - p)^{\beta}, \quad (17)$$

where $\beta = (N-1)(N-2)/2$. The average degree $\langle k \rangle$ is thus

$$\langle k \rangle = \langle 2n \rangle = 2G_0'(1)$$
$$= p(N-1)(N-2) \approx pN^2. \quad (18)$$

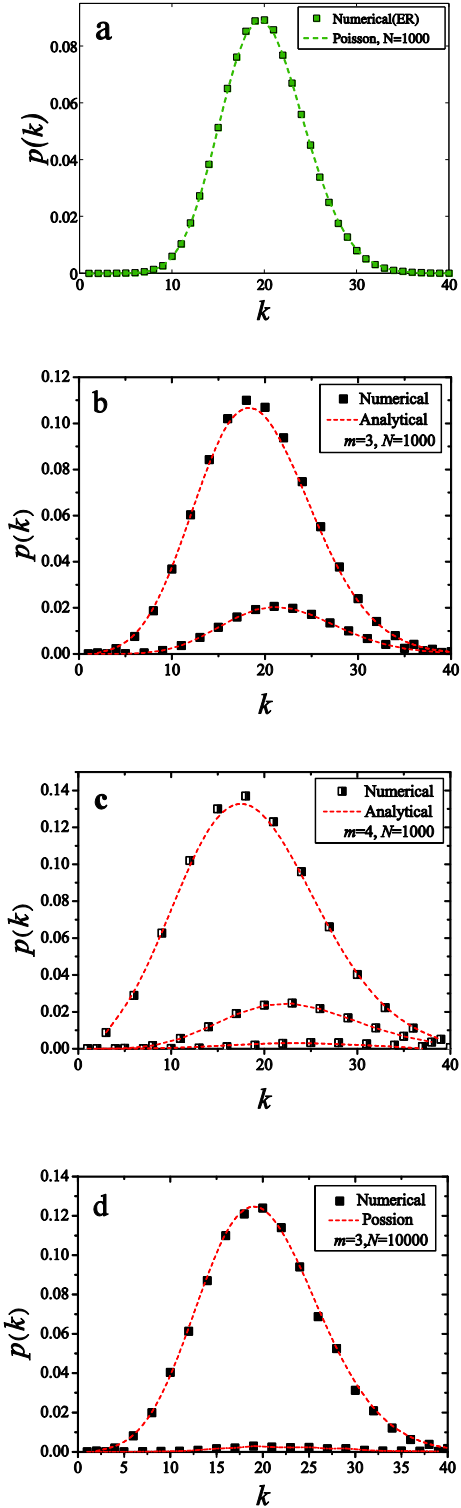

FIG. 3: (Color online) Analytical and numerical results of the degree distribution of RCNs: (a) $m = 2$, $N = 1000$; (b) $m = 3$, $N = 1000$; (c) $m = 4$, $N = 1000$; and (d) $m = 3$, $N = 10000$. We have used different $p$'s for different RCNs so that their average degrees are all approximately 20.

TABLE I: Average degrees of clique networks.

|  | $m=2$ | $m=3$ | $m=4$ |
| --- | --- | --- | --- |
| Theoretical value | 20 | 20 | 20 |
| $N=1000$ | 19.98 | 19.74 | 19.66 |
| $N=10000$ | 19.998 | 19.973 | 19.967 |

Similarly, for $m = 4$, the corresponding clique generating function is

$$G_0(z) = \sum_{n=0}^{\gamma} \binom{\gamma}{n} \cdot p^n (1-p)^{\gamma-n} z^n = (pz + 1 - p)^{\gamma}, \quad (19)$$

where $\gamma = (N-1)(N-2)(N-3)/6$, and the average degree $\langle k \rangle$ is

$$\langle k \rangle = \langle 3n \rangle = 3G_0'(1)$$
$$= \tfrac{1}{2} p(N-1)(N-2)(N-3) \approx \tfrac{1}{2} pN^3. \quad (20)$$

The theoretical and numerical values of the average degrees are given in Table I, and they agree well with each other.

## IV. CLUSTERING COEFFICIENT

One of the most important parameters for complex networks is the clustering coefficient. In this subsection, we present a simulation investigation of the clustering coefficient of random clique networks. The clustering coefficient $c_a$ of a node $a$ is defined as the ratio between the number of edges $e$ among the $k$ neighbors of node $a$ and its maximum possible value, $k_a(k_a - 1)/2$:

$$c_a = \frac{2e_a}{k_a(k_a-1)} \quad (21)$$

The clustering coefficient $C$ of the network is then given by the average of the clustering coefficients of all the nodes:

$$C = \frac{1}{N} \sum_{a \in N} c_a \quad (22)$$

For an ER network, its clustering coefficient is the same as the connection probability p, i.e.,

$$C_{ER} = p = \frac{\langle k \rangle}{N} \quad (23)$$

because the probability of any two nearest neighbors of a given node in the ER network being connected is equal to the probability that two randomly selected nodes are connected.

The clustering coefficients of clique networks with $m = 3$ and 4 contain more information than those of the ER network. We first studied how they varied with the average degree $\langle k \rangle$. In Fig. 4, we plot the clustering coefficients $C$ as a function of $\lg\langle k \rangle$ for different clique orders $m$. For the ER network, we can see from Eq. (23) that $C$ is an exponential function of $\lg\langle k \rangle$. However, for 3- and 4-clique networks, as $\lg\langle k \rangle$ increases, the clustering coefficients $C$, rather than increasing monotonically, first increase and then decrease after reaching a maximum value $\sim 0.5$; $C$ starts to increase again as the network becomes increasingly dense and tends to the value of the ER network for a large $\langle k \rangle$ limit.

The non-monotonic behavior of the relation between $C$ and $\langle k \rangle$ can be understood as follows. When $\langle k \rangle$ is small (i.e.

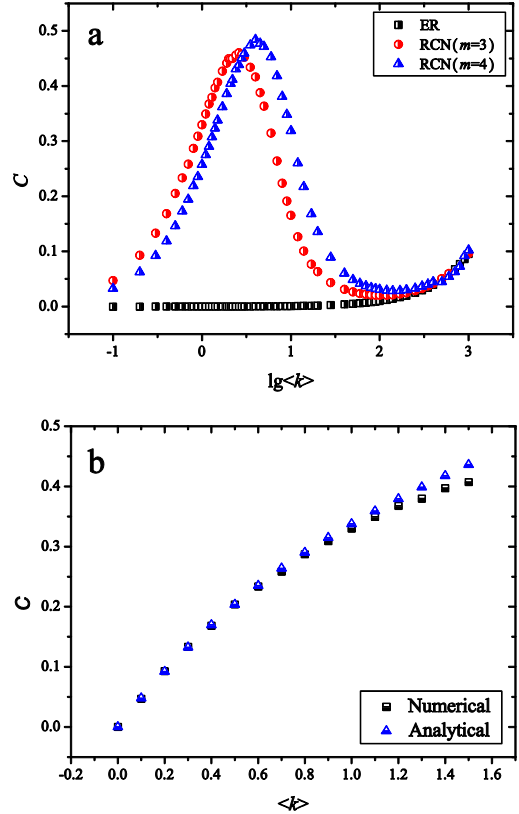

FIG. 4: (Color online) Clustering coefficient $C$ of RCNs ($N = 10000$) as a function of their average degree $\langle k \rangle$. In (a), the results for $m = 2$, $m = 3$, and $m = 4$ are all shown; in (b), both the analytical and numerical results for $m = 3$ in the small $\langle k \rangle$ limit are given.

$p$ is small), most cliques are isolated from each other; thus, the initial rise of $C$ is induced by the increase of the total clique number. As $p$ increases, clusters composed of multiple cliques appear, i.e., there exist nodes shared by multiple cliques. It is easy to show that the clustering coefficients of such nodes are smaller than those for the nodes of isolated cliques. Therefore, the increase of $p$ now has two opposite influences on $C$: on the one hand, it induces the increase of the clique number and would thus increase $C$; on the other hand, the increase of $p$ also increases the fraction of nodes shared by multiple cliques and would thus reduce $C$. It is reasonable to expect that at a certain $p$, the latter effect counteracts the former one, and thus, $C$ decreases when $\langle k \rangle$ increases. The rise of $C$ in the final stage where $\langle k \rangle$ is very large is because the networks are now highly connected, and thus, behave similar to the ER network.

When $p$ is small, the relation between $C$ and $\langle k \rangle$ can be determined analytically. We assume that for small $p$, the network is mainly composed of isolated cliques and clusters with two or three connected cliques. We start with the clusters having three cliques. Given that $N$ is large, we consider only the first two categories shown in Fig. 2, and the total numbers of these two categories are given in Eq. (4) and Eq. (5), respectively. The number of clusters of two connected cliques is then

$$N_{C2} = N_{C2}^T - (2N_1 + 3N_2),  \quad (24)$$

where the configurations of two cliques sharing one edge have been ignored; the number of isolated cliques is

$$N_{iso} = N_{iso}^T - 2N_{C2} - 3(N_1 + 3N_2) ,. \quad (25)$$

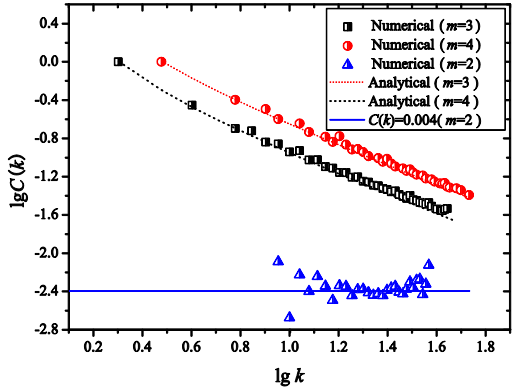

FIG. 5: (Color online) Clustering spectrum of $m = 3$ and $m = 4$ networks with size $N = 5000$ and average degree $\langle k \rangle \simeq 20$. Also plotted is the clustering spectrum of the ER network ($m = 2$), which is expected to be independent of $k$ ($C(k) = \langle k \rangle /N = 0.004$)

Combining Eqs. (4), (5), (24), and (25) with Eqs. (22) and (18), we thus obtain

$$C = \frac{1}{2}\langle k \rangle - \frac{5}{24}\langle k \rangle^2 + \frac{11}{240}\langle k \rangle^3 . \quad (26)$$

We can see from Fig. 4(b) that this analytical approximation agrees well with the corresponding numerical results when $\langle k \rangle$ is small.

## V. CLUSTERING SPECTRUM

The hierarchical structure of complex networks is characterized by their clustering spectrum, which is defined as follows:

$$C(k) = \frac{1}{NP(k)} \sum_{a=1}^{N} C_a \times \delta_{kk_a}, \quad (27)$$

where $P(k)$ is the degree distribution function, $\delta(x)$ is the Kroneckers delta function, and $k_a$ and $c_a$ denote, respectively, the degree and clustering coefficient of node $a$. For networks without a hierarchical structure, such as the ER network, $C(k)$ is independent of $k$[10, 14]. However, for random clique networks, our simulation results show that $C(k) \propto k^{-\alpha}$ with $\alpha \approx 1.01$ for $m = 3$ and $\alpha \approx 1.04$ for $m = 4$ (see Fig. 5), indicating that the networks are hierarchical[10].

The power law behavior of $C(k)$ can be understood as follows. First, we consider a random 3-clique network. For $k \ll N$, we ignore the cases of cliques sharing edges, and thus obtain

$$C(k) = \frac{k/2}{\binom{k}{2}} = \frac{1}{k-1} \quad (28)$$

Similarly, for a random 4-clique network, we have

$$C(k) = \frac{k}{\binom{k}{2}} = \frac{2}{k-1} \quad (29)$$

If $k \gg 1$, the above two equations can be approximately expressed as $C(k) \propto k^{-1}$. It can be seen from Fig. 5 that Eqs. (28) and (29) agree well with the numerical results.

## VI. CHARACTERISTIC PATHLENGTH

We present an investigation on the characteristic path length $L$ of random clique networks, i.e., the average of the geodesic lengths $d_{ij}$ over all pairs of nodes:

$$L = \frac{1}{N(N-1)} \sum_{i,j \in N, i \neq j} d_{ij} \quad (30)$$

For the ER network, it has been proved that $L = \lg N / \lg \langle k \rangle$ [6]. Figure 6 shows that the clique networks with $m = 3$ and $m = 4$ and the ER network have the same characteristic path length as far as their node number and average degree are the same. Therefore, the characteristic path length is determined by the formation rule of the networks rather than the clique size.

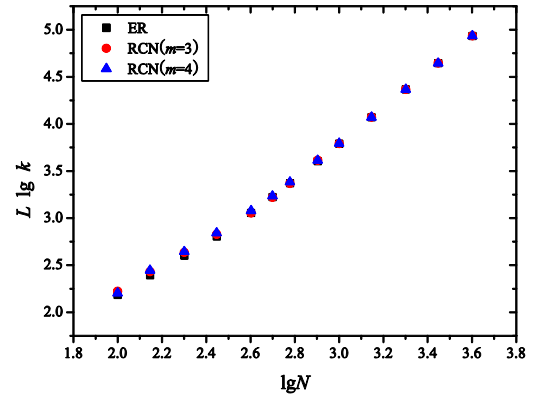

FIG. 6: (Color online) $L \cdot \lg \langle k \rangle$ as a function of $\lg N$ for $m = 2$ (squares), $m = 3$ (circles), and $m = 4$ (triangles) networks.

VII. CONCLUSION

In this article, a random-clique network model was introduced wherein the networks were constructed according to a random selection rule identical to that of the ER model. However, the networks thus constructed showed some distinct properties from ER networks. They are hierarchical networks, possessing large clustering coefficients and a power law clustering spectrum, when their average degrees are small. These networks tend to behave in a manner similar to ER networks when their average degrees are large. Some other unique properties of such networks were also investigated, with the aim of applying them to mimic many real complex networks, such as social networks [25, 28, 29, 30], artificial networks [31, 32], and protein interaction networks [10, 33, 34, 35]. The evolution of the clique clusters and properties of the networks containing cliques of various sizes will be explored in our future work.

**Acknowledgements**

The authors thank Bin Zhou and Changping Yang for helpful discussions. This work was supported by the Chinese Academy of Sciences, the Open Foundation of the State Key Laboratory of Theoretical Physics (Grant No.Y3KF321CJ1), and the National Natural Science Foundation of China (Grant No. 10835005).